\documentstyle[12pt,epsf]{article}

\def\eps{\epsilon}
\def\Ord{{\cal O}}
\def\oneloop{{1 \mbox{-} \rm loop}}
\def\tree{{\rm tree}}

\def\Gr{{\rm Gr}}
\def\Tr{{\rm Tr}}
\def\cg{c_\Gamma}
\def\spa#1.#2{\left\langle#1\,#2\right\rangle}
\def\spb#1.#2{\left[#1\,#2\right]}
\def\Split{\mathop{\rm Split}\nolimits}

\def\N{G^n}
\def\F{G^f}

\begin{document}
\topskip 1cm 
\begin{titlepage}

\noindent hep-ph/9810409
\hspace*{\fill}\parbox[t]{4cm}{
EDINBURGH 98/20 \\ 
MSUHEP-81016 \\ 
UCLA/98/TEP/29\\
October 19, 1998}

\vspace{.1cm}

\begin{center}
{\Large\bf The Infrared Behavior of One-Loop Gluon Amplitudes at
Next-to-Next-to-Leading Order} \\
\vspace{1.cm}

{Zvi Bern}\\
\vspace{.2cm}
{\sl Department of Physics and Astronomy\\
University of California at Los Angeles\\
Los Angeles,  CA 90095-1547, USA}\\

\vspace{.8cm}
{Vittorio Del Duca\footnote{On leave of absence from
I.N.F.N., Sezione di Torino, Italy.}}\\
\vspace{.2cm}
{\sl Particle Physics Theory Group\\
Dept. of Physics and Astronomy\\ University of Edinburgh\\
Edinburgh EH9 3JZ, Scotland, UK}\\

\vspace{.5cm}
and \\
\vspace{.5cm}

{Carl R. Schmidt}\\
\vspace{.2cm}
{\sl Department of Physics and Astronomy\\
Michigan State University\\
East Lansing, MI 48824, USA}\\

\vspace{.5cm}

\begin{abstract}

For the case of $n$-jet production at next-to-next-to-leading order in
the QCD coupling, in the infrared divergent corners of phase space
where particles are collinear or soft, one must evaluate 
$(n+1)$-parton final-state one-loop amplitudes through $\Ord(\eps^2)$,
where $\eps$ is the dimensional regularization parameter.  For the
case of gluons, we present to all orders in $\eps$ the required
universal functions which describe the behavior of one-loop amplitudes
in the soft and collinear regions of phase space.  An explicit example
is discussed for three-parton production in multi-Regge kinematics
that has applications to the next-to-leading logarithmic
corrections to the BFKL equation.

\end{abstract}
\end{center}
\vfil

\end{titlepage}

The quest to obtain ever increasing precision in perturbative QCD
requires calculations to higher orders in the QCD coupling,
$\alpha_s$. Over the years, significant effort has been expended on
computing the next-to-leading order (NLO) contributions to multi-jet
rates within perturbative QCD.  An important next step in the endeavor
to obtain higher precision would be the computation of
next-to-next-to-leading-order (NNLO) contributions to multi-jet rates.
As an example, although the NLO contributions to $e^+ e^- \rightarrow
3$ jets have been computed for some time now~\cite{ert,nk}, the NNLO
contributions have yet to be obtained.  A calculation of these
NNLO contributions would be needed to further reduce the theoretical
uncertainty in the determination of $\alpha_s$ from event shape
variables~\cite{schm}.  Considerable effort has also been expended in
the computation of leading and next-to-leading logarithmic
contributions to the BFKL equations for parton evolution at small $x$
\cite{bfkl,ff}.  These two problems are, of course, connected since
the logarithms that are resummed in the BFKL equations also appear at
fixed orders of perturbation theory.  Indeed, the one-loop Lipatov
vertex may be extracted~\cite{dds,dds2} from one-loop five-gluon helicity
amplitudes~\cite{bdk5g}.

In order to compute a cross section at NNLO, three series of amplitudes are
required in the squared matrix elements: {\it a}) tree-level,
one-loop, and two-loop amplitudes
for the production of $n$ particles; {\it b}) 
tree-level and one-loop amplitudes for the
production of $n+1$ particles; {\it c}) tree-level amplitudes for the
production of $n+2$ particles.  For the case of NNLO $e^+ e^-
\rightarrow 3$ jets the five-parton final-state
tree~\cite{ZFiveJetsBorn} amplitudes, as well as four-parton final-state 
one-loop amplitudes exist in both helicity~\cite{bdkZ4} and
squared matrix-element forms~\cite{cgmZ4}, but as we discuss below, in
order to be used in NNLO computations additional terms enter because
of infrared issues. For the required two-loop three-parton final-state
amplitudes no computations exist, as yet. Indeed, no two-loop amplitude
computations exist for cases containing more than a single
kinematic variable, except in the special cases of maximal
supersymmetry~\cite{byr}.

Besides these amplitudes it is also important to have a detailed
understanding of the infrared singularities that arise from
virtual loops and from unresolved real emission when the
momenta of particles become either soft or collinear.  (For hadronic
initial states infrared divergences are also associated with
initial-state parton distribution functions.)  These infrared
divergences show themselves as poles in the dimensional regularization
parameter $\eps=(4-D)/2$.  By the Kinoshita-Lee-Nauenberg
theorem~\cite{kln} the infrared singularities must cancel for
sufficiently inclusive physical quantities.  However, it is only when
the various contributions are combined that the infrared singularities
cancel.

At NLO the structure of the infrared singularities has been extensively
studied.  The singularities occur in a universal way, i.e.
independent of the particular particle production amplitude
considered.  Accordingly, soft singularities have been accounted for by
universal soft functions \cite{bcm,bg}, and collinear singularities by
universal splitting functions \cite{ap}.  A detailed discussion of the
infrared singularities at NLO for $e^+ e^- \rightarrow$ jets may be
found, for example, in ref.~\cite{gg}.

At NNLO the situation is less developed, although some work has
already been performed to illuminate the structure of infrared
divergences.  In particular, in the squared tree-level amplitudes, any
two of the $n+2$ produced particles can be unresolved; accordingly the
ensuing soft singularities, collinear singularities, and mixed
collinear/soft singularities, have been accounted for by double-soft
functions \cite{bg}, double-splitting functions and soft-splitting
functions \cite{cg}, respectively.  Furthermore, the universal
structure of the coefficients of the $1/\eps^4$, $1/\eps^3$ and
$1/\eps^2$ poles has also been determined \cite{cat} for the two-loop
virtual contributions for $n$-particle productions.

In this letter we shall discuss the $(n+1)$-parton final-state case
{\it b}) when the soft or collinear particles are gluons.  In the
interference term between a one-loop amplitude for the production of
$n+1$ particles and its tree-level counterpart any one of the 
produced particles can be unresolved in the final state; the
phase-space integration gives at most an additional double
pole in $\eps$.  Therefore the expansion in $\epsilon$ of the
interference term starts with a $1/\epsilon^4$ pole, from mixed
virtual/real infrared singularities, and in order to evaluate it to
$\Ord(\epsilon^0)$, the $(n+1)$-parton one-loop amplitude needs to
be evaluated to $\Ord(\epsilon^2)$.  (A similar need to evaluate
one-loop amplitudes to higher orders in $\epsilon$ has been previously
noted in NNLO deep inelastic scattering \cite{van} and in the NLL
corrections to the BFKL equation \cite{ffk}.)  For the case of NNLO
corrections to $e^+ e^- \rightarrow 3$ jets, this would be a rather
formidable task given the already non-trivial analytic structure of the
one-loop $e^+ e^- \rightarrow 4$ partons helicity
amplitudes presented in ref.~\cite{bdkZ4} through $\Ord(\eps^0)$.

A much more practical approach is to evaluate the amplitudes to higher
order in $\eps$ only in the infrared-divergent regions of phase-space.
In the collinear and soft regions the amplitudes factorize into sums
of products of $n$-parton final-state amplitudes multiplied by soft or
collinear splitting functions.  (The splitting functions in this
letter are for amplitudes; the Altarelli-Parisi ones are roughly
speaking the squares of these.)  It is these soft or collinear
splitting functions and the $n$-parton final-state one-loop amplitudes
that must be evaluated to higher order in $\eps$. This is a much
simpler task than evaluating the full $(n+1)$-parton final-state 
amplitudes beyond $\Ord(\eps^0)$.

Here we focus on the issue of supplementing one-loop $(n+1)$-parton 
final-state 
amplitudes that are known to $\Ord(\epsilon^0)$ with higher
order in $\eps$ pieces in the soft and collinear regions of phase space.
The one-loop splitting functions have been given through
$\Ord(\epsilon^0)$ \cite{bddkSusy4,bdk3g2q}, and the one-loop soft
functions through $\Ord(\epsilon^0)$ may be extracted from the known
four- \cite{bk} and five-parton \cite{bdk5g,kst,bdk3g2q} one-loop
amplitudes.  Below, we provide the one-loop gluon splitting and soft
functions to all orders in $\epsilon$, leaving the calculational
details and a complete listing of the one-loop splitting and soft
functions, including fermions, to a forthcoming paper.

As an example, we apply the framework outlined above to the one-loop
amplitude for three-parton production in multi-Regge kinematics
\cite{fl}, for which the produced partons are strongly ordered in
rapidity. In NNLO and in next-to-leading-logarithmic corrections to
two-jet scattering, this amplitude appears in an interference term
multiplied by its tree-level counterpart. Because of the rapidity
ordering, phase-space integration does not yield any collinear
singularities; however, the gluon which is intermediate in rapidity
can become soft.  Accordingly, the one-loop amplitude must be determined
to $\Ord(\epsilon^0)$ plus the contribution with the soft intermediate
gluon evaluated to $\Ord(\epsilon)$ \cite{ffk}.  To determine the soft
gluon contribution we use our all orders in $\eps$ determination of
the soft functions together with previous all orders in $\eps$
determinations of the four-gluon amplitudes~\cite{gsb,bddkSusy4,bm}.

We first briefly review properties of $n$-gluon scattering amplitudes,
since we use these below. The tree-level color decomposition is (see
e.g.\ ref.\cite{mpReview} for details and normalizations)
\begin{equation}
M_n^\tree(1,2,\ldots n) =  g^{(n-2)} \sum_{\sigma \in S_n/Z_n}
{\rm Tr}\left( T^{a_{\sigma(1)}} 
T^{a_{\sigma(2)} }\cdots T^{a_{\sigma(n)}} \right)
 m_n^\tree(\sigma(1), \sigma(2),\ldots, \sigma(n)) \,,
\end{equation}
where $S_n/Z_n$ is the set of all permutations, but with cyclic
rotations removed.  We have suppressed the dependence on the particle 
polarizations $\varepsilon_i$
and momenta $k_i$, but label each leg with the index $i$.  The
$T^{a_i}$ are fundamental representation matrices for the Yang-Mills
gauge group $SU(N_c)$, normalized so that ${\rm Tr}(T^aT^b) =
\delta^{ab}$.  The color decomposition of one-loop multi-gluon
amplitudes with adjoint states circulating in the loop is
\cite{bkcolor}
\begin{equation}
M_n^\oneloop(1,2,\ldots n) =  g^n 
\sum_{j=1}^{\lfloor{n/2}\rfloor+1} \sum_{\sigma \in S_n/S_{n;j}}
 \Gr_{n;j}(\sigma) \, m_{n;j}^\oneloop(\sigma(1),\ldots,\sigma(n)) \,,
\label{OneLoopColor}
\end{equation}
where $\lfloor x\rfloor$ denotes the greatest integer less than or
equal to $x$, $\Gr_{n;1}(1) \equiv N_c \, \Tr\bigl(T^{a_1}\cdots
T^{a_n}\bigr)$, $\Gr_{n;j}(1) = \Tr\bigl(T^{a_1}\cdots
T^{a_{j-1}}\bigr) \, \Tr\bigl(T^{a_j}\cdots T^{a_n}\bigr)$ for $j>1$,
and $S_{n;j}$ is the subset of permutations $S_n$ that leaves the
trace structure $\Gr_{n;j}$ invariant. 
It turns out
that at one-loop the $m_{n;j>1}$ can be expressed in terms of
$m_{n;1}^\oneloop$~\cite{bdkReview}, so we need only discuss this case in this
letter.  The amplitudes with fundamental fermions in the loop 
contain only the $m_{n;1}^\oneloop$ color structures and are scaled by 
a relative factor of $1/N_c$.

The behavior of color-ordered one-loop amplitudes as the momenta of
two color adjacent legs becomes collinear, is~\cite{bddkSusy4,bdk3g2q}
\begin{equation}
m_{n;1}^{\oneloop}\ \mathop{\longrightarrow}^{a \parallel b}\
\sum_{\lambda=\pm}  \biggl\{
{\mathop{\rm Split}\nolimits}^{\tree}_{-\lambda}
(a^{\lambda_a},b^{\lambda_b})\,
      m_{n-1;1}^{\oneloop}(\ldots K^\lambda\ldots)
+{\mathop{\rm Split}\nolimits}_{-\lambda}^{\oneloop}
(a^{\lambda_a},b^{\lambda_b})\,
      m_{n-1}^{\tree}(\ldots K^\lambda\ldots) \biggr\} \, ,
\label{OneLoopCollinear}
\end{equation}
where $\lambda$ represents the helicity, 
$m_{n;1}^{\oneloop}$ and $m_n^{\tree}$ are color-decomposed
one-loop and tree sub-amplitudes with a fixed ordering of legs and $a$ and
$b$ are consecutive in the ordering, with $k_a=zK$ and $k_b=(1-z)K$.
The splitting functions in eq.~(\ref{OneLoopCollinear}) have
square-root singularities in the collinear limit. 
For the case of only gluons, the tree splitting functions
splitting into a positive helicity gluon
(with the convention that all particles are outgoing) is 
\begin{eqnarray}
{\mathop{\rm Split}\nolimits}^{\tree}_+(a^+,b^+)
             &=& 0 \,, \hskip 3.6 cm 
{\mathop{\rm Split}\nolimits}^{\tree}_+(a^-,b^-)
             = {-1\over \sqrt{z (1-z)} \spb{a}.{b} } \,, 
      \nonumber\\
{\mathop{\rm Split}\nolimits}^{\tree}_{+}(a^{-},b^{+})
             &=& {z^2\over \sqrt{z (1-z)} \spa{a}.{b} } \,, 
               \hskip 1. cm 
{\mathop{\rm Split}\nolimits}^{\tree}_{+}(a^{+},b^{-})
             = {(1-z)^2\over \sqrt{z (1-z)} \spa{a}.{b} } \,,
\label{TreeSplit}
\end{eqnarray}
where the remaining ones may be obtained by parity.  The spinor inner
products~\cite{SpinorHelicity,mpReview} are $\spa{i}.j =
\langle i^- | j^+\rangle$ and $\spb{i}.j = \langle i^+| j^-\rangle$,
where $|i^{\pm}\rangle$ are massless Weyl spinors of momentum $k_i$,
labeled with the sign of the helicity.  They are antisymmetric, with
norm $|\spa{i}.j| = |\spb{i}.j| = \sqrt{s_{ij}}$, where $s_{ij} =
2k_i\cdot k_j$.

The one-loop splitting functions are,
\begin{eqnarray}
\Split_+^\oneloop(a^-, b^-) &=& (\F+\N) \Split^\tree_+(a^-, b^-)\,, \nonumber\\
\Split_+^\oneloop(a^\pm, b^\mp) &=& \N\,  \Split^\tree_+(a^\pm, b^\mp)\,,
\label{SplitLoop} \\
\Split_+^\oneloop(a^+,b^+) &=& -\F \,{1\over \sqrt{z (1-z)}}  
 {\spb{a}.{b}\over \spa{a}.b^2}\,. \nonumber
\end{eqnarray}
The function $\F$ arises from the `factorizing' contributions
and the function $\N$ arises from the `non-factorizing' ones 
described in ref.~\cite{bc} and are given through $\Ord(\eps^0)$
by~\cite{bddkSusy4,bdk3g2q}
\begin{eqnarray}
\F &=& {1\over 48 \pi^2} 
 \Bigl(1 - {N_{\! f}\over N_c} \Bigr) z (1-z) + \Ord(\eps)\, ,
\label{FNFuncs}\\
\N &=& 
c_\Gamma \Bigl[
- {1 \over \epsilon^2} 
\Bigl( {\mu^2 \over z(1-z) (-s_{ab})}\Bigr)^\epsilon
 + 2 \ln (z) \ln(1-z)
  - {\pi^2 \over 6}\Bigr] + {\cal O}(\epsilon) \, ,\nonumber
\end{eqnarray}
with $N_{\! f}$ the number of quark flavors and
\begin{equation}
c_{\Gamma} = {1\over 
(4\pi)^{2-\epsilon}}\, {\Gamma(1+\epsilon)\,
\Gamma^2(1-\epsilon)\over \Gamma(1-2\epsilon)}\, .\label{cgam}
\end{equation}
As at tree-level, the remaining splitting functions can be obtained 
by parity.  The
explicit values were obtained by taking the limit of five-point
amplitudes; the universality of these functions for an arbitrary
number of legs was proven in ref.~\cite{bc}.  A listing of one-loop
splitting functions through ${\cal O}(\epsilon^0)$ also involving
fermions may be found in refs.~\cite{bddkSusy4,bdk3g2q}.
To ${\cal O}(\epsilon^0)$, these splitting functions are independent 
of the regularization scheme  parameter,
\begin{equation}
\delta_R = \left\{ \begin{array}{ll} 1 & \mbox{HV or CDR scheme},\\
0 & \mbox{FDH or DR scheme}, \end{array}
\right. \, \label{cp}
\end{equation}
where CDR denotes the conventional dimensional regularization scheme,
HV the 't Hooft-Veltman scheme, DR the dimensional reduction scheme,
and FDH the `four-dimensional helicity scheme.  (For further
discussions on scheme choices see refs.~\cite{bk,cst}.)  

We have extended the above results for one-loop gluon splitting, 
as well as similar ones for soft functions, to
all orders in $\epsilon$ in several ways.  The first way is by
following the methods of ref.~\cite{bc} and extending the discussion
to include soft limits, but being careful to keep all contributions to
higher order in $\eps$.  In this method the contributions are divided
into the classes of `factorizing' contributions, that may be obtained
directly from one-loop three-point Feynman diagram calculations and
from `non-factorizing' contributions, that are linked to the 
infrared-singular poles in $\epsilon$.  An important ingredient in this 
construction is that the
set of all possible loop integral functions that may enter into an
amplitude are known functions to all orders in $\eps$.  The method
makes clear the universality of the splitting and soft functions since
it does not rely on the computation of any particular amplitude.

As a second independent method for obtaining the values of the
splitting and soft functions we have computed the amplitudes $gggH$
using the effective $ggH$ coupling~\cite{Dawson} due to a heavy
fermion loop, again being careful to keep all higher order in $\eps$
contributions. (For a discussion of the calculation valid through
$\Ord(\eps^0)$ see ref.~\cite{crs}.)  This is a convenient
amplitude from which to extract the splitting and soft functions since
it involves only four-point kinematics with one massive leg; the
massive leg $H$ ensures that the $gggH$ amplitude has well defined
limits when gluons are collinear or soft.  As a third independent
check we have also verified that the non-factorizing contributions
obtained for the special case of $N=4$ supersymmetric amplitudes
agree with the above determinations.  (The $N=4$ case has no
factorizing contributions and does not provide a check of these.)  The
$N=4$ four- five- and six-point amplitudes have been given to all
orders in $\eps$ in ref.~\cite{DimShift} making it straightforward to
extract the collinear and soft limits in terms of limits of loop
integral functions.

{}From these calculations, 
our results for the all orders in $\eps$ contributions to the
functions (\ref{FNFuncs}) appearing in 
splitting functions are
\begin{eqnarray}
\F &=& {2 \cg \over (3-2\eps) (2-2\eps) (1-2\eps)} 
\Bigl(1 -\eps\delta_R - {N_{\! f}\over N_c} \Bigr) \, 
\Bigl( {\mu^2 \over - s_{ab}} \Bigr)^\eps  z(1-z) \,, 
\label{FNFuncsAllEps} \\
\N &=&  c_{\Gamma}\, 
\left({\mu^2\over -s_{ab}}\right)^{\epsilon}\, {1\over\epsilon^2} 
\left[ -\left({1-z\over z}\right)^\epsilon \Gamma(1-\epsilon)
\Gamma(1+\epsilon) + 2 \sum_{k=1,3,5,...} \epsilon^k \,
{\rm Li}_k\left({-z\over 1-z}\right) \right]\, , \nonumber
\end{eqnarray}
where the polylogarithms are defined as \cite{lewin}
\begin{equation}
\left. \begin{array}{l}
{\rm Li}_1(z)\ = - \ln(1-z) \\ \displaystyle
{\rm Li}_k(z) = \int_0^z {dt\over t}\,{\rm Li}_{k-1}(t) \qquad (k=2,3,\dots)
\end{array}\right\} = \sum_{n=1}^{\infty} {z^n\over n^k}\, .
\label{polys} 
\end{equation}

It is not difficult to verify that
eq.~(\ref{FNFuncsAllEps}) agrees with eq.~(\ref{FNFuncs})
through $\Ord(\eps^0)$.  Although
not obvious, the expression for $\N$ in eq.~(\ref{FNFuncsAllEps}) is
symmetric in $z\leftrightarrow (1-z)$; indeed its expansion to ${\cal
O}(\epsilon^2)$ may be written as
\begin{eqnarray}
\N &=&  c_{\Gamma}\, 
\left({\mu^2\over -s_{ab}}\right)^{\epsilon}\,
\Biggl\{ \left[z(1-z)\right]^{-\epsilon}\, \left[-{1\over\epsilon^2} +
\ln{z}\ln(1-z) - {\pi^2\over 6} \right] \nonumber\\ &+& \ln{z}\ln(1-z)  
- 2\epsilon \left[ {\rm Li}_3(z) + {\rm Li}_3(1-z) - 
\zeta(3) \right] \label{nfactexp}\\ 
&+& \epsilon^2 \Biggl[ 
- {1\over 6} \ln{z}\ln(1-z) \ln^2[z(1-z)]
- {2\over 3} \ln^2{z}\ln^2(1-z) \nonumber\\ & & 
 + {\pi^2\over 3} \ln{z}\ln(1-z) - 
{7\over 360} \pi^4 \Biggr] \Biggr\} + {\cal O}(\epsilon^3)\, .\nonumber
\end{eqnarray}

The behavior of one-loop amplitudes in the soft limit is very similar to the
above.  As the momentum $k$ of an external leg becomes soft 
the color-ordered one-loop amplitudes become 
\begin{eqnarray}
\lefteqn{ m_{n;1}^\oneloop (..., a,k^\pm, b,...)|_{k\to 0} =} 
\label{OneLoopSoft}\\ 
& & {\rm Soft}^{\tree}(a,k^\pm, b)\, m_{n-1;1}^\oneloop(..., a, b,...) 
+ {\rm Soft}^\oneloop (a,k^\pm, b)\, m_{n-1}^{\tree}(..., a, b,...)\,, 
\nonumber
\end{eqnarray}
with the tree-level soft functions
\begin{equation}
{\rm Soft}^{\tree}(a,k^+,b) = {\left\langle a\,b\right\rangle
 \over \left\langle a\,k\right\rangle \left\langle k\,b\right\rangle}\,,
\hskip 2 cm 
{\rm Soft}^{\tree}(a,k^-,b) = {-[ a\,b]
\over [a\, k] [k\, b]}\,.
\end{equation}
%
Following analogous methods as for the collinear case, we have computed
the one-loop gluon soft function to all orders of $\epsilon$, 
with the result, 
\begin{equation}
{\rm Soft}^{\oneloop}(a,k^\pm,b) = - {\rm Soft}^{\tree}(a,k^\pm,b)\,
c_{\Gamma}\, {1\over\epsilon^2}\,
\left({\mu^2(-s_{ab})\over (-s_{ak})(-s_{kb})}\right)^{\epsilon}\, 
{\pi\epsilon\over \sin(\pi\epsilon)}\, .\label{soft}
\end{equation}
The soft function (\ref{soft}) does not depend on $N_{\! f}$ or $\delta_R$
and through $\Ord(\epsilon^2)$ it is
\begin{equation}
{\rm Soft}^{\oneloop}(a,k^\pm,b) = - {\rm Soft}^{\tree}(a,k^\pm,b)\,
c_{\Gamma}\, \left({\mu^2(-s_{ab})\over (-s_{ak})(-s_{kb})}
\right)^{\epsilon}\, \left({1\over\epsilon^2} + {\pi^2\over 6}
+ {7\pi^4\over 360}  \eps^2 \right)
+ {\cal O}(\epsilon^3)\, .\label{softexp}
\end{equation}
Through $\Ord(\epsilon^0)$ this agrees with the results that may be
extracted from four- \cite{bk} and five-parton
\cite{bdk5g,kst,bdk3g2q} one-loop amplitudes that are known through
$\Ord(\eps^0)$.

We now apply these results for one-loop splitting (\ref{SplitLoop})
and soft (\ref{soft}) functions to the case of three-gluon production 
in multi-Regge kinematics.  To do so, 
we also need the exact four-gluon one-loop amplitude through 
$\Ord(\eps)$.  In fact,
this is known exactly to all orders of
$\epsilon$.  From the string-inspired decomposition
of the (unrenormalized) one-loop four-gluon 
sub-amplitude \cite{bdk3g2q,bdkReview}, we write
\begin{equation}
m_{4;1}^\oneloop = A_4^g + \left(4- {N_{\! f} \over N_c}\right) A_4^f + 
\left(1- {N_{\! f}\over N_c}\right) A_4^s\, ,\label{stri}
\end{equation}
where $A_4^g$, $-A_4^f$, and $A_4^s$ are the contributions from
an $N=4$ supersymmetric multiplet, an $N=1$ chiral multiplet, and 
a complex scalar, respectively.  We can also write
\begin{equation}
A_4^x = c_{\Gamma}\, m_4^\tree\, V^x\, , \qquad\qquad x=g,f,s\,
,\label{dec}
\end{equation}
with $m_4^\tree$ the corresponding tree-level subamplitude.
The functions $V^f$ and $V^s$ depend on the helicity configuration and
can be extracted to all orders in $\eps$ from ref.~\cite{bm} by taking 
the massless limit.  For configurations 
of type $m_{4;1}^\oneloop(1^-,2^-,3^+,4^+)$ this yields,
\begin{eqnarray}
V^f &=& - \tilde{I}_2(s_{23}) - \epsilon\, {s_{23}\over s_{13}}\, 
\tilde{I}_4^{D=6-2\epsilon}\, ,\label{vf}\\
V^s &=& 2\ \left[ \left(1 - \epsilon {s_{23}\over s_{12}}\right)\,
\tilde{I}_2^{D=6-2\epsilon}(s_{23})
+\ {s_{23}\over s_{12}} \epsilon (1-\epsilon) \tilde{I}_4^{D=8-2\epsilon}
\right]\, ,\nonumber
\end{eqnarray}
while for configurations of type $m_{4;1}^\oneloop(1^-,2^+,3^-,4^+)$ we have
\begin{eqnarray}
V^f &=& \left[{s_{23}\over s_{13}}\, \tilde{I}_2(s_{12}) +
{s_{12}\over s_{13}}\, \tilde{I}_2(s_{23}) - {s_{12}s_{23}\over
s_{13}^2}\,(1-\epsilon) \tilde{I}_4^{D=6-2\epsilon}\right]\, ,\nonumber\\
V^s &=& 2\ \Biggl[ - {s_{12}s_{23}(s_{12}-s_{23})\over s_{13}^3}\,
\epsilon\, \left(\tilde{I}_3^{D=6-2\epsilon}(s_{23}) -
\tilde{I}_3^{D=6-2\epsilon}(s_{12})\right) \label{vg}\\ &-&
{s_{12}s_{23}\over s_{13}^3}\, (s_{12}\tilde{I}_2(s_{23})+ s_{23}\tilde{I}_2
(s_{12})) - {1\over s_{13}}\, \left(s_{12}\tilde{I}_2^{D=6-2\epsilon}
(s_{23}) + s_{23}\tilde{I}_2^{D=6-2\epsilon}(s_{12})\right) \nonumber\\ &+& 
{s_{12}s_{23}\over s_{13}^2}\, \epsilon\, \left(\tilde{I}_2^{D=6-2\epsilon}
(s_{23}) + \tilde{I}_2^{D=6-2\epsilon}(s_{12})\right) -
{s_{12}s_{23}\over s_{13}^2}\, \left(\tilde{I}_3^{D=6-2\epsilon}
(s_{23}) + \tilde{I}_3^{D=6-2\epsilon}(s_{12})\right) \nonumber\\ &+&
{s_{12}^2s_{23}^2\over s_{13}^4}\, \tilde{I}_4^{D=6-2\epsilon}\,
+ {s_{12}s_{23}\over s_{13}^2}\, \epsilon (1-\epsilon)\,
\tilde{I}_4^{D=8-2\epsilon}\Biggr]\, ,\nonumber
\end{eqnarray}
with
\begin{eqnarray}
\tilde{I}_2(s) &=& \left({\mu^2\over - s}\right)^{\epsilon}\, {1\over
\epsilon (1-2\epsilon)}\, , \nonumber \\ \tilde{I}_2^{D=6-2\epsilon}(s) &=&
\left({\mu^2\over - s}\right)^{\epsilon}\,
{1\over 2\epsilon (1-2\epsilon)(3-2\epsilon)}\, ,\nonumber \\ 
\tilde{I}_3^{D=4-2\epsilon}(s) &=& -
\left({\mu^2\over - s}\right)^{\epsilon}\,
{1\over\epsilon^2}\, ,\label{is}\\
\tilde{I}_3^{D=6-2\epsilon}(s) &=&
\left({\mu^2\over - s}\right)^{\epsilon}\,
{1\over 2\epsilon (1-2\epsilon)(1-\epsilon)}\, ,\nonumber\\
\tilde{I}_4^{D=4-2\epsilon} &=&
{2\over\epsilon^2} \left[\left({\mu^2\over s_{23}}\right)^{\epsilon}\, 
_{2}F_1\left(-\epsilon, -\epsilon; 1-\epsilon; 1 + {s_{23}\over s_{12}}\right)
+ \left({\mu^2\over s_{12}}\right)^{\epsilon}\, _{2}F_1\left(-\epsilon, 
-\epsilon; 1-\epsilon; 1 + {s_{12}\over s_{23}}\right)
\right]\, ,\nonumber\\
\tilde{I}_4^{D=6-2\epsilon} &=& - {1\over 2 (1-2\epsilon)}
\left[ \tilde{I}_4^{D=4-2\epsilon} + 2 \tilde{I}_3^{D=4-2\epsilon}(s_{23})
+ 2 \tilde{I}_3^{D=4-2\epsilon}(s_{12}) \right]\, ,\nonumber\\
\tilde{I}_4^{D=8-2\epsilon} &=& - {1\over 2 (3-2\epsilon)}
\left[ {s_{12}s_{23}\over s_{13}^2} \tilde{I}_4^{D=6-2\epsilon} + 
2 {s_{23}\over s_{13}} \tilde{I}_3^{D=6-2\epsilon}(s_{23})
+ 2 {s_{12}\over s_{13}}\tilde{I}_3^{D=6-2\epsilon}(s_{12}) \right]\,
.\nonumber
\end{eqnarray}
The functions $\tilde I_n$ are scalar loop integrals in the indicated 
dimension scaled by prefactors so as to make them dimensionless.
The function $V^g$ obtained from the $N=4$
multiplet~\cite{gsb,bddkSusy4,DimShift} has the same functional form
for either helicity configuration. To all orders in $\eps$ it is 
\begin{equation}
V^g = - \tilde{I}_4^{D=4-2\epsilon} - \epsilon\, \delta_R\, V^s
\, .\label{vertica} 
\end{equation}
Any partial amplitude of the type $m_{4:3}^\oneloop$ may then be
obtained from sums of permutations of the $m_{4;1}^\oneloop$ \cite{bkcolor}.

We next need the dispersive part of this amplitude in the high-energy
limit, $s\gg t$.  The leading color orderings of the sub-amplitudes of
type $m_{4;1}^\oneloop$ are \cite{dds} $(A^-,A'^+,B'^+,B^-)$,
$(A^-,B'^+,B^-,A'^+)$, $(A^-,A'^+,B^-,B'^+)$, and
$(A^-,B^-,B'^+,A'^+)$ where we take the Mandelstam variables to be $s
= s_{AB}$, $t = s_{BB'}$, and $u= s_{AB'}$. In eqs.~(\ref{vf}) 
and (\ref{vertica}) orderings of type
$m_{4;1}^\oneloop(-,-,+,+)$ occur with $s_{12}\rightarrow s$ and
$s_{23}\rightarrow t$, while in eqs.~(\ref{vg}) and (\ref{vertica})
orderings of type $m_{4;1}^\oneloop(-,+,-,+)$ occur with $s_{12}\rightarrow u$
and $s_{23}\rightarrow t$ or $s_{12}\rightarrow t$ and
$s_{23}\rightarrow u$.  However, for the second helicity
configuration, the functions in eqs.~(\ref{vg}) and (\ref{vertica})
are symmetric under the exchange of $s_{12}$ and $s_{23}$.  Thus we
can limit the analysis to one ordering for each type.

Using the usual prescription $\ln(t) = \ln(-t) -i\pi$ (for $t<0$), we have
\begin{equation}
{\rm Re}\, \left({\mu^2\over t}\right)^{\epsilon} =
\left({\mu^2\over -t}\right)^{\epsilon}\, \cos(\pi\epsilon)\, .\label{presc}
\end{equation}
In the high-energy limit, $s \gg -t$,
\begin{eqnarray}
_{2}F_1\left(-\epsilon, -\epsilon; 1-\epsilon; 1 + {t\over s_{12}}\right)
&=& \Gamma(1-\epsilon)\,\Gamma(1+\epsilon)\, +\ 
\Ord\left({t\over s_{12}}\right)
\nonumber\\ &=& {\pi\epsilon\over \sin(\pi\epsilon)}\, +\ 
\Ord\left({t\over s_{12}}\right)\, ,\label{highf}\\
\left({\mu^2\over s_{12}}\right)^{\epsilon}\, _{2}F_1\left(-\epsilon, 
-\epsilon; 1-\epsilon; 1 + {s_{12}\over t}\right) &=& 
\left({\mu^2\over -t}\right)^{\epsilon}\, 
_{2}F_1\left(-\epsilon, 1; 1-\epsilon; 1 + {t\over s_{12}}\right)
\nonumber\\ &=& - \left({\mu^2\over -t}\right)^{\epsilon}\, \epsilon\,
\left[\psi(1) - \psi(-\epsilon) + \ln{s_{12}\over -t}\right] +\
\Ord\left({t\over s_{12}}\right)\, .\nonumber
\end{eqnarray}
with $s_{12}\rightarrow s$ or $s_{12}\rightarrow u$. Since $u=-s-t$, for either 
choice of $s_{12}$
the dispersive parts are the same to the required accuracy, and we can take 
$s_{12}\rightarrow s$ in eqs.~(\ref{highf}). Substituting 
eq.~(\ref{presc}) and (\ref{highf}) in eq.~(\ref{is}), and using the identity
\begin{equation}
- \pi {\cos(\pi\epsilon)\over \sin(\pi\epsilon)}\, =\
\psi(1+\epsilon) - \psi(-\epsilon)\, ,\label{iden}
\end{equation}
we obtain
\begin{equation}
{\rm Disp}\,\tilde{I}_4^{D=4-2\epsilon} = 
- {2\over\epsilon}\, \left({\mu^2\over -t}
\right)^{\epsilon}\, \left[\psi(1+\epsilon) - 2\psi(-\epsilon) +
\psi(1) + \ln{s\over -t}\right] +\ 
\Ord\left({t\over s}\right)\, .\label{nfour}
\end{equation}
In addition, in the high-energy limit
both eq.~(\ref{vf}) with $s_{12}\rightarrow s$ and 
$s_{23}\rightarrow t$, and eq.~(\ref{vg}) with $s_{12}\rightarrow u$ 
and $s_{23}\rightarrow t$ reduce to
\begin{eqnarray}
V^f &=& -\ \left({\mu^2\over - t}\right)^{\epsilon}\, {1\over
\epsilon (1-2\epsilon)} +\ \Ord\left({t\over s}\right)\, ,\label{vfhigh}\\
V^s &=& \left({\mu^2\over - t}\right)^{\epsilon}\, {1\over
\epsilon (1-2\epsilon)(3-2\epsilon)} +\ 
  \Ord\left({t\over s}\right)\, ,\nonumber
\end{eqnarray}
while the function $V^g$, eq.~(\ref{vertica}), is obtained from
eq.~(\ref{nfour}) and (\ref{vfhigh}).

Using eqs.~(\ref{stri}), (\ref{dec}), (\ref{vertica}), (\ref{nfour})
and (\ref{vfhigh}), and the fact that the proportionality factor 
between each tree-level subamplitude and its one-loop
correction is the same for all color orderings, we obtain the
unrenormalized four-gluon one-loop amplitude in the
high energy limit to all orders in $\epsilon$,
\begin{eqnarray}
\lefteqn{ {\rm Disp}\, M_4^\oneloop(A^-,A'^+,B'^+,B^-) =  
M_4^{\tree}(A^-,A'^+,B'^+,B^-)\, g^2\, c_{\Gamma}\, 
\left({\mu^2\over - t}\right)^{\epsilon}\,
{1\over \epsilon (1-2\epsilon)} } \label{alleps}\\ &\times&
\Biggl\{ N_c\, \left[2(1-2\epsilon) \left(\psi(1+\epsilon) - 
2\psi(-\epsilon) + \psi(1) + \ln{s\over -t}\right) + {1 - 
\delta_R\epsilon\over 3-2\epsilon}
- 4\right] + {2(1-\epsilon)\over 3-2\epsilon} N_{\! f} \Biggr\}\,. 
\nonumber
\end{eqnarray}
Following the methods of ref.~\cite{dds}, we can then extract
the unrenormalized one-loop correction to the helicity-conserving 
vertex, to NLL accuracy,
\begin{eqnarray}
\lefteqn{{{\rm Disp}\, C^{gg(1)}_{-+}(-p_a,p_{a'})\over 
C^{gg(0)}_{-+}(-p_a,p_{a'})} =
{{\rm Disp}\, C^{gg(1)}_{-+}(-p_b,p_{b'})\over
C^{gg(0)}_{-+}(-p_b,p_{b'})} = c_{\Gamma}\, 
\left({\mu^2\over - t}\right)^{\epsilon}\, {1\over \epsilon (1-2\epsilon)} }
\label{allvert}\\ &\times& \Biggl\{ N_c\, \left[(1-2\epsilon) 
\left[\psi(1+\epsilon) 
- 2\psi(-\epsilon) + \psi(1) \right] + {1 - \delta_R\epsilon
\over 2(3-2\epsilon)}
- 2\right] + {1-\epsilon\over 3-2\epsilon} N_{\! f} \Biggr\}\, .\nonumber
\end{eqnarray}
Eq.~(\ref{allvert}) is valid to all orders in $\epsilon$ for
$\delta_R=0$ and 1; it agrees with the one-loop correction to the 
helicity-conserving vertex computed in ref.~\cite{dds} to $\Ord(\epsilon^0)$
and with the one computed in ref.~\cite{fl,ffl} to all orders in
$\epsilon$, for $\delta_R=1$.

In an inclusive high-energy two-jet cross section at NNLO and in the
NLL corrections to the BFKL kernel, the five-gluon one-loop amplitude
is multiplied by the corresponding tree-level amplitude
with the intermediate gluon $k$ integrated over its phase space. 
Assuming multi-Regge kinematics, the only infrared divergence that 
can arise is in the soft limit for the intermediate gluon, $k\to 0$.
It gives a single pole in $\epsilon$. Thus, in order to
generate correctly all the finite terms in the squared amplitude, the
five-gluon one-loop amplitude must be computed exactly to
$\Ord(\epsilon^0)$, and must be augmented by the $\Ord(\epsilon)$
corrections in the soft limit for the intermediate gluon.  To achieve
that, we need the dispersive part of eq.~(\ref{OneLoopSoft}) in the
physical region, $s_{ab} > 0$, $s_{ak} > 0$, $s_{kb} > 0$. 
(The other leading color orderings yield the same result.)  Using
eq.~(\ref{soft}), (\ref{presc}), and the identity (\ref{iden}), we can
write the dispersive part of the soft function to all orders in
$\epsilon$ as,
\begin{equation}
{\rm Disp\,\, Soft}^{\oneloop}(a,k^\pm,b) =
-{\rm Soft}^{\tree}(a,k^\pm,b)\,
{c_{\Gamma}\over\epsilon^2}\,
\left({\mu^2\,s_{ab}\over s_{ak}\,s_{kb}}\right)^{\epsilon}\,
\left[1 + \epsilon \psi(1-\epsilon) - \epsilon \psi(1+\epsilon)\right]\,
 .\label{dispers}
\end{equation}
In addition, using the strong rapidity ordering,
\begin{equation}
y_a \gg y \gg y_b; \qquad |k_{a\perp}|\simeq|k_{\perp}|
\simeq|k_{b\perp}|\, ,\label{mrk}
\end{equation}
the mass-shell condition for the intermediate gluon gives
\begin{equation}
s_{ab} = {s_{ak}\, s_{kb}\over |k_{\perp}|^2}\, .\label{mass}
\end{equation}
Subsequently, eq.~(\ref{dispers}) becomes,
\begin{equation}
{\rm Disp\,\, Soft}^{\oneloop}(a,k^\pm,b) =
-{\rm Soft}^{\tree}(a,k^\pm,b)\,
{c_{\Gamma}\over\epsilon^2}\, \left({\mu^2\over 
|k_{\perp}|^2}\right)^{\epsilon}\, \left[1 + \epsilon \psi(1-\epsilon) 
- \epsilon \psi(1+\epsilon)\right]\,
 .\label{hedisp}
\end{equation}
The soft limit for the tree-level five-gluon sub-amplitudes,
\begin{equation}
m_5^\tree(A^-,A'^+,k^\pm,B'^+,B^-) = {\rm Soft}^{\tree}
(A',k^\pm,B')\, m_4^\tree(A^-,A'^+,B'^+,B^-)\, ,\label{fatto}
\end{equation}
holds for arbitrary kinematics.
Thus, the unrenormalized five-gluon one-loop 
amplitude in the multi-Regge kinematics, and in the soft
limit for the intermediate gluon and to all orders in $\epsilon$, is 
obtained by using eq.~(\ref{OneLoopSoft}), with the four-gluon one-loop 
amplitude~(\ref{alleps}), the loop soft function~(\ref{hedisp}),
and eq.~(\ref{fatto}), yielding
\begin{eqnarray}
\lefteqn{ {\rm Disp}\, M_5^\oneloop(A^-,A'^+,k^\pm,B'^+,B^-)\bigr|_{k\to 0} =  
 g^2\, c_{\Gamma}\, M_5^{\tree}(A^-,A'^+,k^\pm,B'^+,B^-)\bigr|_{k\to 0}\, } 
\nonumber\\ &\times& \Biggl[ \left({\mu^2\over - t}\right)^{\epsilon}
\Biggl\{ N_c\, \Biggl[-{4\over\epsilon^2} + {2\over\epsilon}\,
\left(\psi(1+\epsilon) - 2\psi(1-\epsilon) + \psi(1) + \ln{s\over -t}\right) 
\label{softlim}\\ &+& {1\over \epsilon (1-2\epsilon)}\,\left({1 - 
\delta_R\epsilon\over 3-2\epsilon} - 4\right)\Biggr] + 
{2(1-\epsilon)\over \epsilon (1-2\epsilon)(3-2\epsilon)} N_{\! f} \Biggr\} 
\nonumber\\ &-& N_c \left({\mu^2\over 
|k_{\perp}|^2}\right)^{\epsilon}\, {1\over\epsilon^2}\,
\left[1 + \epsilon \psi(1-\epsilon) - \epsilon \psi(1+\epsilon)\right]
\Biggr]\, .\nonumber
\end{eqnarray}
To $\Ord(\epsilon)$, eq.~(\ref{softlim}) reads
\begin{eqnarray}
\lefteqn{ {\rm Disp}\, M_5^\oneloop(A^-,A'^+,k^\pm,B'^+,B^-)\bigr|_{k\to 0} =  
 g^2\, c_{\Gamma}\, M_5^{\tree}(A^-,A'^+,k^\pm,B'^+,B^-)\bigr|_{k\to 0}\, } 
\nonumber\\ &\times& \Biggl\{ \left({\mu^2\over - t}\right)^{\epsilon}
\Biggl[ N_c\, \Biggl(-{4\over\epsilon^2} + {2\over\epsilon}\ln{s\over -t}
+ \pi^2 - {64\over 9} - {\delta_R\over 3} + 2\zeta(3)\epsilon 
- {380\over 27}\epsilon - {8\over 9}\delta_R\epsilon \Biggr) 
\label{softexpan}\\
&-& {\beta_0\over \epsilon} + N_{\! f} \left({10\over 9} + {56\over 27}
\epsilon\right) \Biggr] - N_c \left({\mu^2\over 
|k_{\perp}|^2}\right)^{\epsilon}\, \left({1\over\epsilon^2} - 
{\pi^2\over 3}\right) \Bigg\} + \Ord(\epsilon^2)\, ,\nonumber
\end{eqnarray}
with $\beta_0 = (11N_c-2N_{\! f})/3$. We have checked that
eq.~(\ref{softexpan}) agrees to $\Ord(\epsilon^0)$ with 
the five-gluon one-loop amplitude, with strong rapidity ordering and 
in the soft limit for the intermediate gluon \cite{bdk5g,dds2}.  
The above result may be used to verify the virtual next-to-leading
log corrections to the Lipatov vertex for use in the BFKL equation as
is done in ref.~\cite{dds2}.

The same type of analysis may be applied more generally to the problem
of NNLO QCD corrections.
The one-loop gluon splitting and soft functions that we have
presented here are valid to all orders in the dimensional
regularization parameter, $\eps$.  This allows them to be used in NNLO
calculations with infrared singular phase space where terms of up to
two powers in $\eps$ are necessary.  Previous explicit determinations
of the one-loop collinear splitting functions~\cite{bddkSusy4,bdk3g2q}
were not performed to the required order in $\eps$.  A systematic
discussion of the soft and collinear splitting functions and further
calculational details, including the case of external fermions, will 
be presented elsewhere.  In particular, these functions can be used to 
aid in the computation of NNLO contributions to $e^+ e^- \rightarrow 3$ 
jets once all the matrix elements are available.  However, much more 
remains to be done before this is realized.

\vskip .2 cm 

This work was supported by the US Department of Energy under grant
DE-FG03-91ER40662, by the US National Science Foundation under grant
is PHY-9722144 and by the EU Fourth Framework Programme {\em Training and
Mobility of Researchers}, Network {\em Quantum Chromodynamics and the 
Deep Structure of Elementary Particles}, contract FMRX-CT98-0194 
(DG 12 - MIHT). The work of V.D.D. and C.R.S was also supported by
NATO Collaborative Research Grant CRG-950176.


\end{document}